\documentstyle[twoside,ijmpa,epsfig]{article}
\def\qed{\hbox{${\vcenter{\vbox{                        
   \hrule height 0.4pt\hbox{\vrule width 0.4pt height 6pt
   \kern5pt\vrule width 0.4pt}\hrule height 0.4pt}}}$}}

\def\bsc{{\sc a\kern-6.4pt\sc a\kern-6.4pt\sc a}}       
\def\bflatex{\bf L\kern-.30em\raise.3ex\hbox{\bsc}\kern-.14em
T\kern-.1667em\lower.7ex\hbox{E}\kern-.125em X}

\begin{document}
\runninghead{Acceptance dependence of fluctuation
$\ldots$} {$\ldots$ in particle multiplicity}
\normalsize\textlineskip
\thispagestyle{empty}
\setcounter{page}{1}

\copyrightheading{}                     

\vspace*{0.88truein}

\fpage{1}
\centerline{\bf 
ACCEPTANCE DEPENDENCE OF FLUCTUATION IN PARTICLE MULTIPLICITY}

\vspace*{0.37truein}

\centerline{\footnotesize D.~P.~Mahapatra,
  B.~Mohanty\footnote{Corresponding author, e-mail: bedanga@iopb.res.in
    }
~and S.~C.~Phatak
}

\vspace*{0.015truein}

\centerline{\footnotesize\it Institute of Physics}
\baselineskip=10pt
\centerline{\footnotesize\it Bhubaneswar, 751-005 India}
\vspace*{0.225truein}


\vspace*{0.21truein}

\abstracts{The effect of limiting the acceptance in rapidity on event-by-event
multiplicity fluctuations in nucleus-nucleus collisions has been 
investigated. Our analysis shows that the multiplicity fluctuations 
decrease when the rapidity acceptance is decreased. We explain this trend by 
assuming that the  probability distribution of the particles in the 
smaller acceptance window follows binomial distribution. 
Following a simple statistical analysis we conclude that the
event-by-event multiplicity fluctuations for full acceptance are likely 
to be larger than those observed in the experiments, since the 
experiments usually have detectors with limited acceptance. We
discuss the application of our model to simulated data generated 
using VENUS, a
widely used event generator in heavy-ion collisions. We also discuss
the results from our calculations in presence of dynamical
fluctuations and possible observation of these in the actual data.}{}{}

\vspace*{1pt}\textlineskip      
\vspace*{-0.5pt}

\section{Introduction}

The analysis of individual ultra-relativistic nucleus-nucleus collision
events has now become feasible as the number of particles produced in
such events is large~\cite{qm99,qm01}. In last few years, 
the subject of event-by-event analysis in general and fluctuations in
various  
observables in particular has attracted significant attention\cite{fluc}. 
This attention has partly been motivated by the fact that, if the 
evolution of the nuclear matter during the collision passes close to 
the expected tri-critical point in the nuclear phase
diagram\cite{tricrit}, the fluctuations in physical observables will
be affected. One of the quantities studied in details is the 
the fluctuation in the numbers of charged particles\cite{NA49} and
photons\cite{WA98}. Particularly, the quantity of interest is $ W =
\frac{ <\Delta N^2 > } { < N > } $. It was observed that this quantity 
is close to $2$ for charged particles as well as photons and
it is also found to depend on the centrality (or impact parameter)
of collision.  If one assumes that the particle production is purely a
statistical process, determined by a Poisson distribution, 
$ W $ is expected to be unity. Thus the departure of 
$W$ from unity was attributed to the dynamics of the
collision and attempts were made to explain it from different
models. As it turns out, two extreme models, the thermal
model\cite{thermal} and the initial-state interaction model ( the
so-called wounded nucleon model )\cite{wounded} are able to explain
this number equally well~\cite{heisel}. 
This may mean that $ W $ is not the right quantity to distinguish 
between such diverse models or that the origin of this departure is 
not in the dynamics but somewhere else.

One must note that the experimental data on charge particle or photon
multiplicities is available in the restricted region of the total
phase space available for the reaction. 
For example, the experimental data in WA98
experiment is restricted to the pseudo-rapidity between $2.9 < \eta
<  4.2$ for photons and $2.35 < \eta < 3.75$ for charged particles.
So far, the variation  of $W$ as a function of the phase space has not
been investigated. Of course, a priori, one does not expect a strong
variation of $W$ with the phase space. In fact, the models, 
such as the thermal models, imply phase space independence of
$W$. On the other hand, a naive argument would imply increase in the
fluctuations when the phase space is reduced. Thus, the investigation
of the dependence of $W$  
on the phase space region chosen for the detection is of importance
since this may affect the
conclusions drawn regarding the physical processes taking place during
the reaction. In fact, preliminary analysis of WA98
photon and charge particle data\cite{private} indicates that $W$ decreases 
when the $\eta$-acceptance is reduced. A similar trend is also
observed in simulations ( see later for the details ). Such a behavior of $W$
needs to be explained. Another possibility is that the interesting physics,
such as QGP formation, may be restricted to a limited region of the
phase space. That is, in a nucleus-nucleus collision, instead of the
whole system, a part of it may undergo phase transition. In that case,
the nature of fluctuations may depend 
on the size of the phase space in which the interesting physics takes
place. It would therefore be useful to investigate the phase
space dependence of the experimental data.
However a recent theoretical calculation in a very different context
to what is presented here, has emphasized the importance
of studying acceptance dependence. They have studied the effect of
acceptance on conserved charged fluctuations~\cite{charge_fluc}.

In the present work we investigate the aspects mentioned in the
preceding paragraphs. In particular, we study the variation
of $W$ as the rapidity acceptance is changed. 
As a concrete example, consider that the experimental measurement
consists of, say, the number of charged particles 
detected in a given range of 
rapidity [$\eta_1$,$\eta_2$]. Further we select the events having 
certain range of of transverse energy $\Delta E_T$, which in some sense 
corresponds to a certain range of impact parameter ($\Delta b$).
In several experiments\cite{NA49,WA98} it has been established that
the distribution of the number of charged particles ( or photons ) in
these events follow Gaussian distribution 
closely. For these data, one can calculate $W$ defined earlier. The
calculations yield a value close to 2 and a number of physical
explanations have been offered for this
value\cite{thermal,wounded}. 
These explanations do not depend on the
range of the rapidity window and therefore one would expect that the
measured $W$ should be independent of it. Below we shall show that
this expectation does not hold and it is important to carefully
analyze the acceptance dependence of
fluctuations before concluding on presence or absence of dynamical
fluctuation in the experimental data.

\section{Statistical fluctuation}

Let us consider the consequences of limiting the acceptance on the
fluctuations. We shall assume that the probability of finding a
particle in the rapidity window $\Delta \eta$ is given by binomial
or normal distribution. For this we shall consider that the particles are
emitted in a full rapidity range [$\eta_1$,$\eta_2$] and in an
experiment one actually detects the particles in a smaller rapidity
window $\Delta \eta$. We can then compute $W$ for the smaller 
acceptance and determine its dependence on the rapidity window $\Delta \eta$.  

\subsection{Binomial Distribution}

Consider a situation in which the distribution of particles in a rapidity
window $\Delta \eta$ is decided by a binomial distribution. That is,
the probability that a particle will be in $\Delta \eta$ is $p$ and the
distribution is purely statistical. Then the probability of finding $n$
particles in $\Delta \eta$ out of total $N$ particles in
[$\eta_1$,$\eta_2$] is given by

\begin{equation}
P(n,N) = \frac{N ! }{ n ! (N - n) !} p^{n} ( 1 - p)^{N-n}  
\end{equation}

For the binomial distribution, the first and second moments are 
$<n(N)> = p N$ and $<n^2(N)> = Np(1-p) + N^{2}p^{2}$. 
Then  fluctuation in $n$ for a fixed value of $N$ is  given as,

\begin{equation}
W(n,N) = \frac{<n^2(N)> - <n(N)>^2}{<n(N)>} = 1- p~; 0\leq p \leq1
\end{equation}

Let us now assume that the event-by-event distribution of 
particles detected in
[$\eta_1$,$\eta_2$] is given according to a normal distribution 
with average number $N_0$ and variance $\sigma$ ( Actually, one can
consider any other
distribution whose first two moments are $N_0$ and $\sigma$
respectively ). Now we need to
compute $<n>$ and $<n^2>$ when $N$'s are distributed according
to the  distribution defined above. A straight forward calculation
results in, 

\begin{equation}
<n> = p <N> = p N_0
\end{equation}
and 
\begin{equation}
<n^2> =<N> p(1-p) + <N^{2}>p^{2}.
\end{equation} 
With this  
\begin{equation}
W = 1 - p + p \frac{\sigma^2} {N_0} = 1 -p + p W_0 
\end{equation}

Where $W_0$ = ${\sigma^{2}}$/$N_{0}$. For Poisson distribution,
$W_0$ is unity and as mentioned earlier, its value is close to 2 in
experiments.

The expression of $W$ obtained above is interesting. What it tells is
that if $W_0$ for the rapidity range  [$\eta_1$,$\eta_2$] is
unity, as in case of Poisson distribution or when $ \sigma^2 = <N>$
for normal distribution, $W$ for the rapidity range $\Delta \eta$ is
also unity. On the other hand, if $W_0$ is larger than unity, $W$ lies 
between unity and $W_0$ and $W$ approaches unity in the limit of small 
$p$. Thus, contrary to the expectations, the fluctuations at the
smaller rapidity window become smaller. We can, in fact, turn this argument
around. In any present day heavy-ion experiment, one never 
has complete  acceptance of the phase space for a given particle
species. Therefore, if the
particles falling  in the limited acceptance of an experiment are
following binomial distribution, the `actual' value of $W_0$ for the
experiment should be larger than the measured $W$ for a
limited acceptance $\Delta \eta$. Unfortunately, it is not possible to 
use this argument to extrapolate and obtain $W_0$ for full acceptance. 

\subsection{Gaussian Distribution}

Now consider a situation in which the probability of
observing $n$ particles in $\Delta \eta$ ( out of $N$ which are
emitted in [$\eta_1$,$\eta_2$]) is given by Gaussian distribution of
the form,

\begin{equation}
P(n) = \frac{1}{\sqrt{2\pi{\sigma_{n}}^{2}}} e^{-\frac{(n-p
    N)^{2}}{2{\sigma_{n}}^{2}}} 
\end{equation}

where $p = n/N$,
with $<n(N)> = p N$ and $<n^2(N)> = {\sigma_{n}}^{2} +
{p}^{2} N^{2}$ \\

If we assume that the particles detected in
[$\eta_1$,$\eta_2$] are distributed according to a Gaussian distribution 
with average number $N_0$ and variance $\sigma$, we will need to
compute $<n>$ and $<n^2>$ when $N$'s are distributed according
to the Gaussian distribution. The simple calculation yields,

$<n> = p <N> = p N_0$ and $<n^2> = {\sigma_{n}}^{2} +
{p}^{2} <N^{2}> $. \\ 
This gives :

\begin{equation}
W = \frac{{\sigma_{n}}^{2}} {p N_0} + p W_0 
\end{equation}

This shows that depending
on the value of $p$ for a given $W_0$ the fluctuation in a
smaller acceptance ($W$) can be higher as well as lower than that of $W_0$.

\section{Application to simulated data}

Ideally, one should test the validity of our proposal with the
experimental data. This is being done. In this work we demonstrate
the application of the above ideas to the simulated $Pb+Pb$ events at
$158 \cdot A GeV$ generated from VENUS~4.12 event
generator~\cite{venus}. 45K VENUS events were generated between impact
parameter $0-12~fm$. In order to minimize the contribution due to
fluctuation from impact parameter, the distributions of $\pi^{0}$,
$\pi^{-}$ and $\pi^{+}$ were chosen 
within narrow $5\%$ bins of cross section 
of the minimum bias transverse energy ($E_{T}$). 
It must be mentioned that since impact
parameter is not directly measurable in experiment, observable like
transverse energy or forward energy can be used to define the
centrality of the reaction. 
A highly central event corresponds to lower  impact parameter event 
and leads to higher transverse energy production.
To determine the transverse energy of the reaction, 
we have done a fast simulation in which we calculate transverse energy
from VENUS by taking the resolution factors
for the hadronic and electromagnetic energy of a realistic calorimeter
as used in WA80 and WA98 experiment into account~\cite{et_cal}.
The resolution of transverse electromagnetic energy was taken to be
$17.9\%/\sqrt E$ and that for hadronic energy was $46.1\%/\sqrt E$,
where $E$ is expressed in GeV.

\begin{figure}[h]
\begin{center}
\includegraphics[scale=0.6]{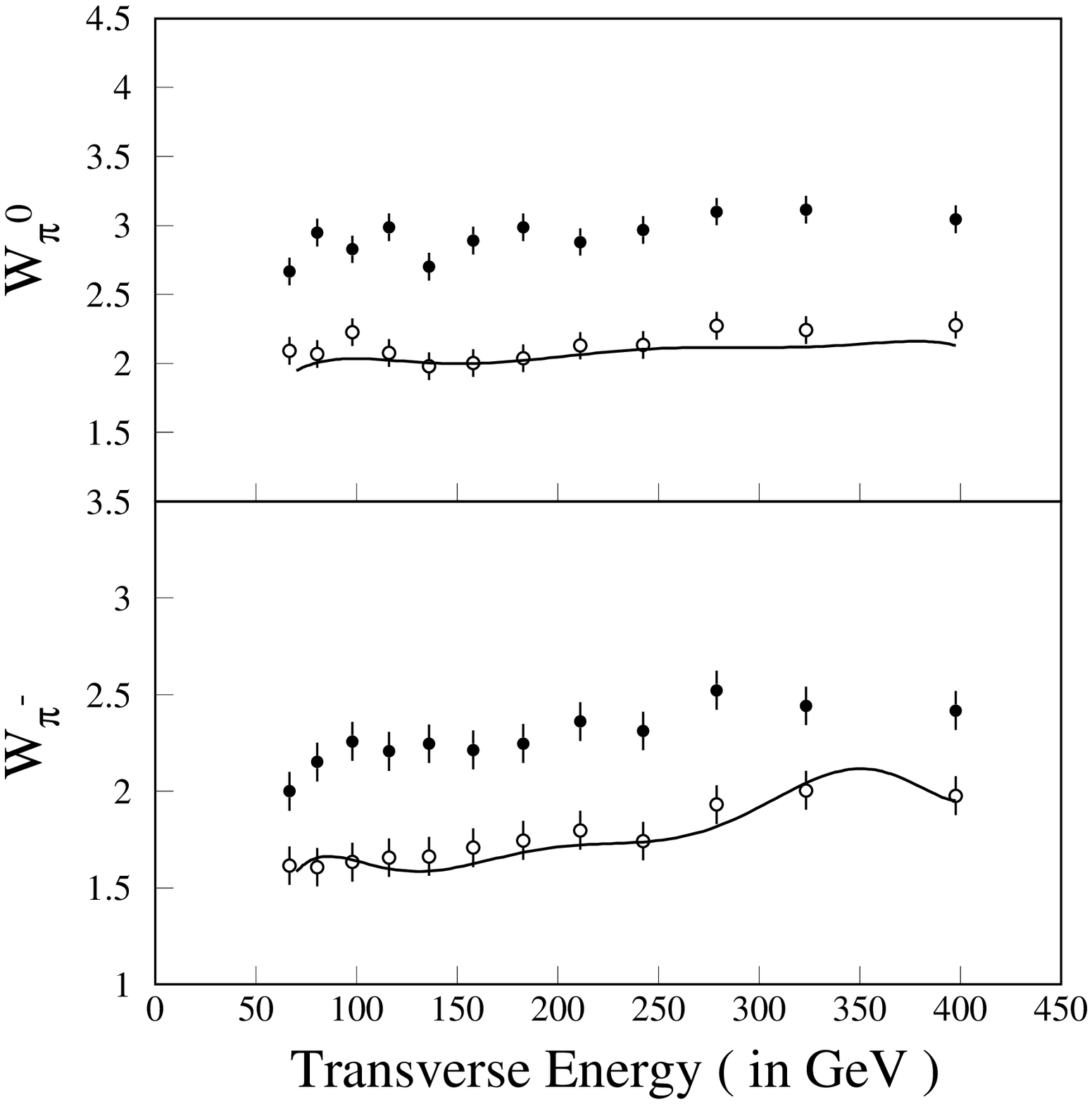}
\caption{ Fluctuation for $\pi^{0}$ and $\pi^{-}$ 
obtained from VENUS for $2$ unit $\Delta \eta$ coverage (solid points)
and for $1.0$ unit $\Delta \eta$ coverage (open points). The model 
calculations are shown as solid line. The statistical errors are within
the symbol size. The results for $\pi^{+}$ are similar to those
obtained for $\pi^{-}$.
}
\label{venus_binomial}
\end{center}
\end{figure}

Event-by-event particle multiplicity distribution is obtained within the 
rapidity range [$\eta_1$,$\eta_2$] $= [2.0,4.0]$, 
for different $5\%$ bins in centrality ($0-5\%$, $5-10\%$ $\cdots$ $55-60\%$).
These distributions are then fitted to Gaussians and fit
parameters mean ($N_{0}$) and standard deviation ($\sigma$) are
obtained. Fluctuation $W_0 = \sigma^{2}/N_{0}$ is then
calculated. We then select one unit $\Delta \eta$ window, and obtain
the multiplicity distributions, the Gaussian fit parameters of these
distributions are used to obtain fluctuations $W$.
We find that $W < W_0$ for 
all the transverse energy bins. The centrality dependence is 
shown in Figure~\ref{venus_binomial}.  

 Using the value of $W_{0}$ calculated for [$\eta_1$,$\eta_2$]  
 in Eqn.~5 we obtain the new $W_{model}$  for $1.0$ unit
 $\Delta \eta$. These are shown in Figure~\ref{venus_binomial}, 
 for $\pi^{0}$ and $\pi^{-}$, $\pi^{+}$ shows similar trend as for $\pi^{-}$.
 It can be seen that the model calculations (solid lines) match very well with
 values of fluctuation from VENUS in smaller acceptance (open circles). The
 statistical errors are shown in the figure.  Similar procedure has
 been followed for the actual experimental data, to verify the above discussed
 models and distributions. The preliminary results of the
 calculation for actual data is seen to follow a similar trend\cite{private}.

\section{Presence of dynamical fluctuation}

An interesting possibility is that an exotic physical process is
restricted to a region of phase space. An example is the formation of
quark matter in a small volume in the early phase of the
collision. One can argue that this volume maps into a certain 
( $\eta$, $\phi$ ) region of the phase space~\cite{Bj}. Let us assume that this
region is smaller than the rapidity range [$\eta_1$,$\eta_2$] but
covers the $\Delta \eta$ region of the phase space and leads to an
enhancement in multiplicity compared to a statistical case. Now, the fraction
of particles going into the rapidity region $\Delta \eta$ will depend
on whether an exotic process occurs in the rapidity region $\Delta
\eta$. Let us assume that these probabilities are $p$ and $q$ for
exotic process not occurring and occurring in $\Delta \eta$
respectively. Assuming that fraction $\beta$ of the events have exotic
process occurring in the rapidity region $\Delta \eta$, the expected
value of $W$ can be obtained as follows.

If $P_{1}(n)$ and $P_{2}(n)$ are binomial distributions, with
probabilities $p$ and $q$ respectively. 
Further if $\beta$ is the fraction of 
events out of total events having probability distribution $P_{1}(n)$
then the Probability distribution of all the events is given as,

\begin{equation}
P(n) = \beta P_{1}(n) + (1-\beta) P_{2}(n)
\end{equation}

with
$<n(N)> = \beta p N + (1-\beta) q N$ \\
and $<n^2(N)> = \beta p (1-p)N + (1-\beta) q (1-q)N + (\beta p N +
(1-\beta) q N)^{2}$

The fluctuation can be calculated to be 
\begin{equation}
W = 1 - \frac{\beta p^{2} + (1-\beta) q^{2}}{\beta p + (1-\beta) q} +
W_0 (\beta p + (1-\beta) q) 
\end{equation}

One can see that in various limiting cases the fluctuation $W$
approaches the pure statistical case discussed in the previous section :
\begin{itemize}

\item Case~1 : $\beta$~=~$1 \Rightarrow W = (1-p) + p W_{0}$
\item Case~2 : $\beta$~=~$0 \Rightarrow W = (1-q) + q W_{0}$
\item Case~3 : $p=q \Rightarrow W$ = $(1-p) + p W_{0}$

\end{itemize}
            
\begin{figure}[h]
\begin{center}
\includegraphics[scale=0.6]{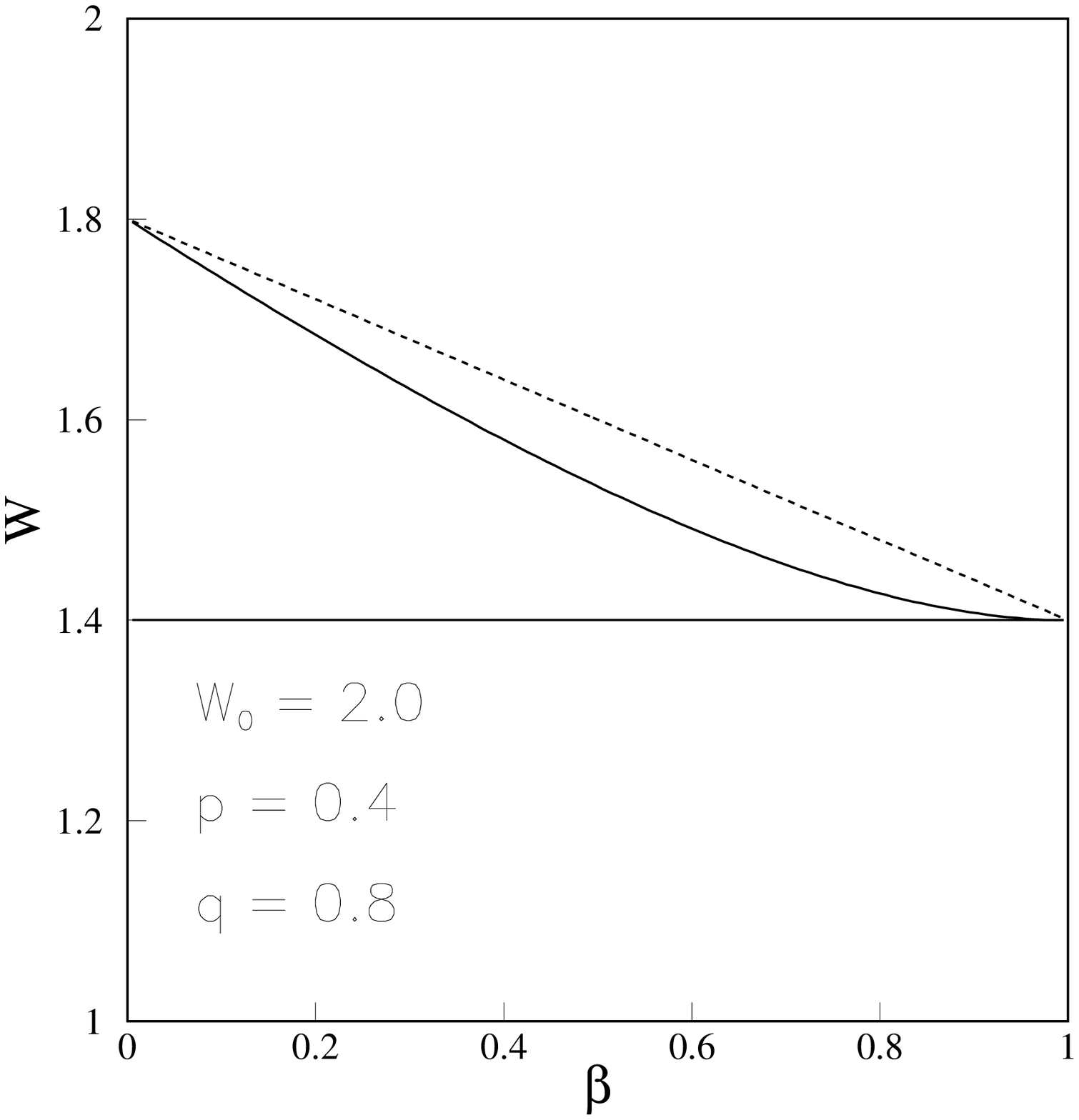}
\caption{ Fluctuation (W) for smaller acceptance as a function
probability of events being statistical type ($\beta$), for 
$q=0.8$. The fluctuation in larger acceptance ($W_0$)
is taken as $2$ and $p = 0.4$. 
}
\label{qgp}
\end{center}
\end{figure}

Figure~\ref{qgp} shows the variation of $W$
with fraction of events having statistical fluctuation, for a
 $W_{0} = 2$ in the full coverage and probability
$p = 0.4$. One can see that, the fluctuation in limited acceptance
is reduced to $1.4$. However as the percentage
of events having dynamical fluctuation ($1-\beta$) increases, for 
$q = 0.8$, the fluctuation in limited acceptance also increases (solid 
line).
The differences are clearly visible for events sets having dynamical
fluctuation for more than $10\%$ of events. In actual experiment one
does not know which events have dynamical fluctuation in an ensemble
of events collected over a period of data taking. So analysis of data
will yield a fixed $W$ following Eqn.~5 as :

\begin{equation}
W = 1 - \bar{p} + \bar{p} \frac{\sigma^2} {N_0} 
  = 1 - \bar{p} + \bar{p} W_0 
\end{equation}

Where $\bar{p} = \beta p + (1-\beta) q$. The variation of this $W$
with $\beta$ is also shown as dashed line in Figure~2. It shows a 
linear increase with decrease in $\beta$. The differences between the
values of $W$ from Eqn's 9 and 10 are beyond statistical errors shown
in figure~\ref{venus_binomial} for certain range of $\beta$. 
With increased statistics one should
in principle observe this difference. If presence of dynamical
fluctuations has
an impact parameter dependence, i.e, if probability of QGP-type fluctuations
increases with increases in centrality of the reaction due to increase
in energy density, then one should study the centrality dependence of
$W$. The $\beta$ values may be replaced by centrality parameter or 
impact parameter in figure~\ref{qgp}, and a deviation from linear 
dependence would indicate
presence of dynamical fluctuations. It must be mentioned that as the value 
of $q$ decreases the difference between the two curves also reduces and
they match with each other as $\beta$ approaches  $0$ and $1$. These set
the constraints in detecting the dynamical fluctuations.

Another possibility of detecting dynamical fluctuations through our
model from actual experimental data, where one will obtain a single value of 
fluctuation $W$, is through construction of mixed events. In order to
make a conclusion regarding the 
presence of dynamical fluctuations, one has to compare this value
to what is expected from a pure statistical process. For this 
a set of events has to be  
constructed from data, so that they preserve all the inherent detector
related fluctuations and broad global features, like total
multiplicity in an event but remove the local dynamical fluctuations.
Such events are referred to as mixed events or mixed data and
techniques for constructing these exist ( see for example
Ref.~\cite{WA98-DCC} ). 
A comparative study of fluctuations obtained from data and mixed
events constructed from data will help arriving at proper conclusions
regarding the presence of dynamical fluctuations as discussed above.

\section{Summary}

In the present work we have investigated the dependence of the
event-by-event multiplicity fluctuations on the acceptance of the 
detector. Unlike the naive expectations 
the fluctuations in the data 
generated from VENUS event generator\cite{venus} {\it decrease} when 
the acceptance is decreased. Similar tendency is also observed in the
preliminary results from WA98 data\cite{private}. 
We find that the trend of the simulated data
is reproduced by assuming that the probability distribution in the
smaller rapidity window follows binomial distribution. 
From this observation, one may infer that the event-by-event
fluctuations in the  experiment, when the full rapidity acceptance is
included, may be larger than the experimentally observed fluctuations
since the experiments are necessarily restricted to smaller rapidity
acceptance. One needs to keep this mind when one is attempting to
explain the fluctuations from theoretical models. 
We have also investigated the possibility of detecting the dynamical
fluctuations if some special dynamics is restricted to a small phase
space.


\begin{thebibliography}{99}


\bibitem{qm99}      Proceedings of {\it Quark Matter '99},
                    Nucl. Phys A {\bf 661} (1999).

\bibitem{qm01}      Proceedings of {\it Quark Matter '2001},
                    To appear in Nucl. Phys A.

\bibitem{fluc} L. van Hove, Phys. Lett. {\bf B118} (1982) 138;
                   E.V. Shuryak, Phys. Lett. {\bf B423} (1998) 9;
                   M. Gazdzicki and S. Mrowczynski, Z. Phys. {\bf C57}
                   (1992) 127; S. Mrowczynski, Phys. Rev. C {\bf 57} (1998)
                   1518; Phys. Lett. {\bf B430} (1998) 9;
                   M.Asakawa, U.Heinz, and B.M{\"u}ller, Phys. Rev. Lett.
                   {\bf 85}, 2072 (2000);S. Jeon and V. Koch
                   Phys. Rev. Lett. 
                    {\bf 83} (1999) 5435; S. Jeon and V. Koch,
                    Phys. Rev. Lett. 
                    {\bf 85}, 2076 (2000).

\bibitem{tricrit}    M. Stephanov, K. Rajagopal and E. Shuryak, Phys. Rev.
                  Lett. {\bf 81} (1998) 4816; Phys. Rev. D {\bf 61}
                  (1999) 114028.


\bibitem{NA49}      H. Appelshauser et al., (NA49 Collaboration),
                    Phys. Lett. {\bf B459} (1999) 679; G. Roland,
                    Nucl. Phys. A{\bf 638} (1998) 91c.

\bibitem{WA98}      T.K.~Nayak et al., (WA98 Collaboration),
                    nucl-ex/0103007.

\bibitem{thermal}    G. Bertsch, Phys. Rev. Lett. {\bf 72} (1994) 2349.
                  (1999) 114028.


\bibitem{wounded}      A. Bialas et al.,
                    Nucl. Phys. B{\bf 111} (1976) 461.


\bibitem{heisel}   G. Baym and H. Heiselberg, Phys. Lett. {\bf B469}
                   (1999) 7; M. Stephanov et al., hep-ph/9903292. 

\bibitem{private}    M.M.~Aggarwal et al., (WA98 Collaboration)
                     e-print: nucl-ex/0108029.

\bibitem{charge_fluc} M. Bleicher, S. Jeon, and V. Koch, 
                      Phys. Rev. C{\bf 62} (2000) 061902(R).  
     

\bibitem{venus}     K. Werner, Phys. Rep. {\bf C232} (1993) 87.

\bibitem{et_cal}       T.~C.~Awes et al., Nucl. Instrum. Methods Phys. Res.
                     Sect. A {\bf 279}, 497 (1989).


\bibitem{Bj}        J.D. Bjorken, Phys. Rev. {\bf D27}, (1983) 140.

\bibitem{WA98-DCC}    M.M.~Aggarwal et al., (WA98 Collaboration)
                     Phys. Rev. {\bf C64}, (2001) 011901(R).

\end{thebibliography}
\end{document}